\documentclass[aip,graphicx,reprint]{revtex4-1}

\usepackage[sort&compress]{natbib}
\usepackage{amssymb}
\usepackage{booktabs} 
\usepackage[T1]{fontenc}
\usepackage[ngerman,english]{babel}
\usepackage{amsfonts}
\usepackage{amsmath}
\usepackage{array}
\usepackage{epsf}
\usepackage{epsfig}
\usepackage{float}
\usepackage[version=3]{mhchem}
\usepackage{textcomp}
\usepackage{graphics}
\usepackage{epstopdf}
\usepackage{placeins}
\usepackage{footnote}
\usepackage{bbm}
\usepackage{enumerate}

\usepackage{color}
\newcommand{\cc}[1]{\textcolor{black}{#1}} 
\newcommand{\Fig}[1]{Fig.~\ref{#1}}
\newcommand{\eq}[1]{Eq.~(\ref{#1})}
\newcommand{\OH}{\ce{[Heme\ B-OH]^0}}
\newcommand{\DMSO}{\ce{[Heme\ B-DMSO]^+}}
\newcommand{\HOH}{\ce{[Heme\ B-H_2O]^+}}
\newcommand{\Cl}{\ce{[Heme\ B-Cl]^0}}

\begin{document}

\title{Towards an ab initio theory for metal L-edge soft X-ray spectroscopy of molecular aggregates}
\author{Marie Preu\ss e} \author{Sergey I. Bokarev} \email{sergey.bokarev@uni-rostock.de}
\affiliation{ Institut f\"{u}r Physik, Universit\"{a}t Rostock,
  Albert-Einstein-Str.~23-24, 18059 Rostock, Germany} 
\author{Saadullah G. Aziz}
\affiliation{Chemistry Department, Faculty of Science, King Abdulaziz
  University, 21589 Jeddah, Saudi Arabia}
\author{Oliver K\"{u}hn} \affiliation{ Institut f\"{u}r Physik,
  Universit\"{a}t Rostock, Albert-Einstein-Str.~23-24, 18059 Rostock, Germany}
\date{\today}

\begin{abstract}
  The Frenkel exciton model was adapted to describe X-ray  absorption and resonant inelastic scattering spectra of 
  polynuclear transition metal complexes by means of restricted active
  space self-consistent field method. The proposed approach allows to substantially decrease the requirements to computational resources if compared to a full supermolecular quantum chemical treatment. This holds true in particular in cases where the dipole approximation to the electronic transition charge density can be applied. The computational protocol was applied to the calculation of X-ray spectra of the hemin complex, which forms dimers in aqueous solution. The aggregation effects were found to be comparable to the spectral alterations due to the replacement of the axial ligand by solvent molecules.
\end{abstract}


\maketitle

\section{Introduction}
\label{sec:intro}
Soft X-ray L-edge spectroscopy has become a standard technique to investigate the intricate details of the electronic structure of transition metal compounds.
The most popular variants encompass the X-ray Absorption (XAS) and Resonant Inelastic X-ray Scattering (RIXS) spectroscopies allowing to address the properties of both unoccupied and occupied valence molecular orbitals.~\cite{de_groot_core_2008,milne_recent_2014}
However, the accurate theoretical prediction of L-edge core- and valence-excited electronic states of transition metal compounds often requires to take into account multi-configurational and spin-orbit coupling effects. 
Here, the combination of the Restricted Active Space Self-Consistent Field (RASSCF) method~\cite{malmqvist:1990} and the atomic mean-field integral approximation~\cite{malmqvist:2002} has been proven to be a versatile computational tool.~\cite{josefsson12_3565,bokarev13_083002,suljoti13_9841,atak13_12613,pinjari_2014,engel_2014,bokarev_mn:2015,wernet_2015,grell_2015,pinjari_2016} 
For metal complexes, where the spectroscopically active region is rather localized, the prediction of L-egde spectra requires an active space including  the 2p and all orbitals with notable metal 3d-contributions.
Such a choice corresponds to account for the most important correlation terms as well as dipole allowed transitions. Further, it allows to keep the active space quite compact and the number of considered electronic states of the order of hundreds or a few thousands.

However, the treatment of systems with multiple metal centers goes beyond the current numerical capabilities because of the fast growth of the number of configurations with the size of the active space. 
Moreover, the number of relevant electronic states scales then as tens or hundreds of thousands what hinders the theoretical interpretation of the structure of multicenter systems such as molecular aggregates.
To cope with such situations, in the present article we adopted a strategy known from the theory of excitation energy transfer in molecular aggregates. \cite{may11,schroter15_1} 
Here, the total system is decomposed into its constituent monomers such that local electronic excitations can be clearly defined and the computational effort is substantially reduced.  Such an approach will be particularly justified in cases where the
monomers forming the aggregate are held together by van der Waals forces. 
Considering core-hole excitations, which are rather localized at a particular metal atom, such a separation strategy might even be justified for multiple metal centers within a covalently bound complex.

\begin{figure}[t]
  \includegraphics{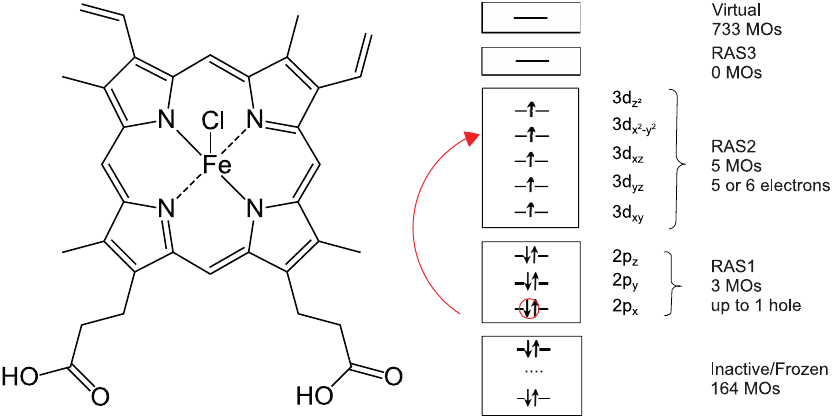}
  \caption{Structure of the hemin, \Cl, molecule  (left) and orbital active space used for calculations (right).
  }
  \label{fig:structure}
\end{figure}

The versatility of this exciton coupling approach is exemplified in the present proof of concept study of the hemin molecule (see~\Fig{fig:structure}).
In addition to its high biological relevance as a constituent of the hemoglobin active center, the hemin complex is interesting due to its aggregation behavior in different solvents.
Staying monomeric in polar solvents like ethanol or DMSO it forms dimers in water solution.~\cite{Villiers_2006}
The effect of aggregation was recently addressed by means of soft X-ray Fe L-edge absorption spectroscopy in transmission (XAS) and partial fluorescence (PFY) modes as well as by off-resonant X-ray emission (XES) and RIXS on the example of DMSO and aqueous solutions.~\cite{atak14_9938,golnak15_3058} 
The general shape of the spectra for both cases was quite similar and the pronounced difference in broadenings for RIXS as well as 1.3~eV energy shift in off-resonant XES was found as an indication of aggregation. These effects were, however, solely attributed to the $\pi-\pi$-stacking,~\cite{golnak15_3058} although ligand $\pi$-orbitals are barely influencing local Fe 2p$\rightarrow$3d transitions measured with L-edge spectroscopy.
Although very recently it was shown that K-edge absorption and emission spectra show features which
can be associated with $\pi$-type interactions,~\cite{golnak15_29000} the interaction upon electronic excitation could be, in general, more complex due to the resonant coupling  of essentially degenerate electronic states of the two monomers.~\cite{may11}
Here, we present a first principles approach capable of quantifying the various coupling-induced effects such as energetic shifts and redistribution of oscillator strengths.
In addition, the labile equilibrium between different species in solution could be a source of spectral changes. That is why in the present paper, we have also studied  the influence of the axial ligand on the X-ray spectra of the monomeric hemin. The following species were investigated: \Cl (original form), \DMSO, \OH, \HOH, including the solvent molecules present in DMSO and aqueous solutions. 

The paper is organized as follows. In Section \ref{sec:theory}, we first present the general theory for the exciton coupling model as adapted to the calculation of first- and second-order X-ray spectra. For the sake of simplicity we restrict ourselves to the case of the dipole approximation. Further, the one- and two-particle approximations will be introduced as a means for restricting the accessible excitation space. The computational setup is detailed in Section \ref{sec:compdet}. In Section \ref{sec:res} we first discuss speciation effects for the monomeric spectra. Subsequently, dimer spectra are provided for different orientations of the two hemin monomers. Section \ref{sec:concl} provides summary and conclusions.

%
\section{Theory}
\label{sec:theory}
\begin{figure}[tb]
  \includegraphics{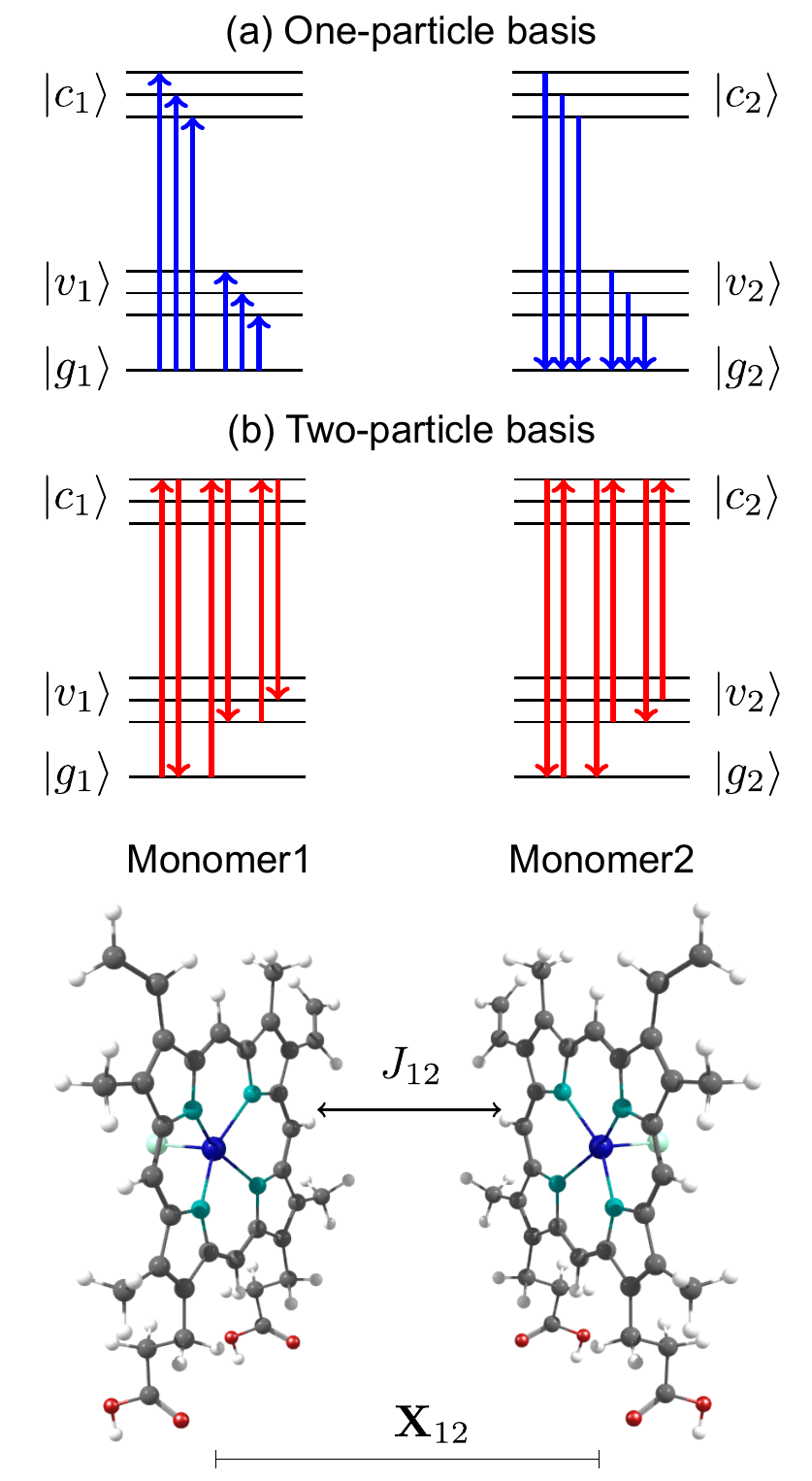}
  \caption{Different choices of excitonic bases and the corresponding transitions included
    in the dimer coupling. For the one-particle basis (a), only transitions from or to the ground state have been
    included (one-particle approximation: OPA). For the two-particle basis (b), de-excitations from an arbitrary core-excited state to any other state are allowed (two-particle approximation: TPA).}
\label{fig:exc}
\end{figure}
In the following we will consider a molecular aggregate and label the monomers by $M$. Let
us denote the electronic states of monomer $M$ by $|A_M\rangle$ with the ground state
being $|A_M\rangle =|g_M\rangle$. Separating the total Hamiltonian according to
\begin{equation}
  \label{eq:htot}
  H_{\rm tot}= \sum_M H_M + \frac{1}{2}\sum_{M,N} V_{MN}
\end{equation}
these states are solutions of the monomeric Schr\"odinger equations
\begin{equation}
  \label{eq:hmono}
  H_M |A_M\rangle = E_{A_M}|A_M\rangle \, .
\end{equation}
In \eq{eq:htot} $V_{MN}$ is the matrix element of the Coulomb operator between the monomer states. Assuming that there
is no wave function overlap between local excitations of the two monomers, the aggregate
wave function can be written in direct (Hartree) product form. This yields the matrix
representation of the aggregate Hamiltonian (assuming frozen nuclei)~\cite{may11}
\begin{eqnarray}
  \label{eq:hmat1}
  H_{\rm tot} &=& \sum_M \sum_{A} E_{A_M} |A_M\rangle \langle A_M | \nonumber\\
              &+& \frac{1}{2} \sum_{MN}
                  \sum_{A,B,C,D}J_{MN}(A_MB_N,C_ND_M) \nonumber\\
&\times&  |A_M \rangle \langle D_M | \otimes |B_N \rangle \langle C_N  |\,. 
\end{eqnarray}
Here, the Coulomb matrix elements $J_{MN}(A_MB_N,C_ND_M) $ have been introduced, which can
be expressed in terms of generalized monomeric total charge densities,
$\mathcal{N}_{A_M,D_M}(\mathbf{r})$,
\begin{equation}
  \label{eq:coulint}
  J_{MN}(A_MB_N,C_ND_M) = \int d\mathbf{r} d\mathbf{r}'\, \frac{\mathcal{N}_{A_M,D_M}(\mathbf{r}) \mathcal{N}_{B_N,C_N}(\mathbf{r}')}{|\mathbf{r}-\mathbf{r}'|}\, .
\end{equation}
\cc{Indices $A,B,C$, and $D$ denote different electronic states of monomers $M$ and $N$.}
Provided that the separation between the monomers, e.g. center to center distance,
$|\mathbf{X}_{MN}|$, is large as compared to the extension of the transition densities and
that the coupling to charge densities can be neglected (e.g. for charge neutral systems),
the transition dipole approximation can be invoked, which gives\cite{may11}
\begin{eqnarray}
  \label{eq:jdip}
  J_{MN}(A_MB_N,C_ND_M) &\approx &\frac{\mathbf{d}_{A_MD_M}\cdot \mathbf{d}_{B_NC_N}}{|\mathbf{X}_{MN}|^3}\nonumber\\
                        & -& 3 \frac{(\mathbf{X}_{MN}\cdot\mathbf{d}_{A_MD_M})(\mathbf{X}_{MN}\cdot \mathbf{d}_{B_NC_N})}{|\mathbf{X}_{MN}|^5}\nonumber\\
                        &&
\end{eqnarray}
where $\mathbf{d}_{A_MD_M}= \langle A_M |{ {\mathbf d}}  |D_M \rangle$ are the transition dipole matrix elements, with
${ {\mathbf d}}$ being the electronic dipole operator. Note that as long as metal-centered transitions are considered, the transition densities are rather localized and the dipole approximation should be valid despite the close proximity of the Heme B planes.

Let us specify the situation to that of a dimer ($M=1,2$) as shown in Fig.~\ref{fig:exc}. 
 The different monomeric state manifolds will be denoted as ground $|g_M\rangle$,
valence-excited, $|v_M\rangle$, and core-excited, $|c_M \rangle$, states.  
Equation (\ref{eq:hmat1}) contains couplings between all possible transitions in the dimer
system.
In the following we will make use of the fact that the actual processes of interest,
namely XAS and RIXS, are of first and second order, respectively, with
respect to the interaction with the external field. This suggests to employ either a one- or two-particle
basis. In the former, states of the type $|a_1g_2\rangle$ and $|g_1a_2\rangle$ ($a=v,c$)
are incorporated, while the latter includes in addition states of type $|a_1,b_2\rangle$
($a,b=v,c$) and is in principle exact for the dimer. Note that this effectively corresponds to a CI-doubles-like
treatment of the composite system with X-ray specific preselection of configurations.
The respective Hamiltonian matrix is readily calculated in terms of the monomeric
excitation energies and the Coulomb integrals in Eq.~\eqref{eq:hmat1}.
To make the calculation computationally feasible, different approximations have been
applied (cf. Fig.~\ref{fig:exc}) to reduce
the size of Hamiltonian matrix as well as the number of terms in RIXS expression (see below):
\begin{enumerate}[(a)]
\item couplings between $g \leftrightarrow v$ transitions (e.g., $J_{12}(v_1 g_2,v_2 g_1)$) and between $g \leftrightarrow v$ and $v \leftrightarrow c$ transitions (e.g., $J_{12}(v_1 c_2,v_2 g_1)$) were neglected, i.e. set to zero in Eq.~(\ref{eq:hmat1}),
\item no coupling between static dipoles as well as between dipoles and transition dipoles have been included, 
\item a pre-selection of core-excited states to construct the basis functions has been applied according to an energy window with a width of
$\pm 5\sigma$ ($\sigma$ is the width of the Gaussian excitation
pulse) around the center of the excitation pulse.
\end{enumerate} 
Note that in contrast to common Frenkel exciton theory not only resonant couplings have been included, but we also partially account for induction and dispersion effects.~\cite{megow_2015}  The use of the one- and two-particle basis together with these conditions will be called one- (OPA) and two-particle approximation (TPA), respectively.

Upon diagonalization of the resulting Hamiltonian matrix, one obtains
eigenstates ($|i\rangle$, $|n\rangle$, and $|f\rangle$ for initial,
intermediate, and final states, respectively) and the respective  transition dipole moments ($\mathbf{d}_{ni}$ and $\mathbf{d}_{fn}$) from
which XAS
\begin{equation}
  \label{eq:SXAS}
  S_{\rm XAS}(E_{\rm{exc}})= \sum_i w(E_i)(E_f - E_i) | \mathbf{ e}_1\cdot \mathbf{d}_{ni}  |^2 \delta(E_{\rm{exc}}-E_f + E_i)
\end{equation}
and RIXS
\begin{eqnarray}
\label{eq:SRIXS}
S_{\rm RIXS}(E_{\rm{exc}},E_{\rm{em}} )& = & 
\sum_i w(E_i) \sum_f \delta(E_{\rm{exc}} + E_{i} - E_{\rm{em}} - E_{f})
\nonumber\\
&\times& \bigg | \sum_n   \frac{(\mathbf{e}_2 \cdot\mathbf{d}_{fn}) (\mathbf{ e}_1\cdot \mathbf{d}_{ni})}{E_i + E_{\rm{exc}} - E_n - i\Gamma_n} \bigg |^2
\end{eqnarray}
spectra can be calculated. Here, $w(E_i)$ is the Boltzmann weight of the corresponding initial state $|i \rangle$ and $\mathbf{e}_1$ ($\mathbf{e}_2$) is the polarization of the incoming (outgoing) light.  The inner sum in Eq.~\eqref{eq:SRIXS} corresponds to the matrix element of the electronic polarizability with respect to the initial and final states.  For the calculation of Partial Fluorescence Yield (PFY)
spectra, the 2D-RIXS spectra have been summed up over the emission energy $E_{\rm{em}}$, yielding
an integrated amplitude over all possible outgoing photon energies in the emission energy
range $[E_{\rm min},E_{\rm max}]$ corresponding to a fixed excitation $E_{\rm{exc}}$:
\begin{equation}
\label{eq:SPFY}
S_{\rm PFY}(E_{\rm{exc}}) = \int_{E_{\rm min}}^{E_{\rm max}} dE_{\rm{em}}\, S_{\rm RIXS}(E_{\rm{exc}},E_{\rm{em}})
\; .
\end{equation}
Note that polarization effects in the XAS and RIXS spectra were taken into account as described in Ref.~\citenum{luo_1996}, assuming a free tumbling of the molecules in solution under the condition that the polarization is detected orthogonal to the incoming beam polarization in laboratory frame.
%
\section{Computational details}
\label{sec:compdet}
%
The energies and transition dipoles for the monomeric Heme B derivatives have been calculated using  geometries, which were
optimized with density functional theory (BLYP functional and the LANL2DZ basis set for
iron and a 6-311+G(d) basis set  for all other elements) using Gaussian 09.~\cite{frisch09_}

Subsequently, the electronic wave function has been
determined via a RASSCF calculation as implemented in the Molcas 8.0 program package
\cite{Molcas} using an ANO-RCC triple zeta basis set with [21s15p10d6f4g2h]/(6s5p3d2f1g)
contraction for iron, [8s4p3d1f]/(2s1p) for hydrogen, [14s9p4d]/(3s2p1d) for carbon and
oxygen, [14s\allowbreak9p\allowbreak4d\allowbreak3f]/(4s3p2d1f) for nitrogen,
[17s12p5d4f]/(5s4p2d1f) for chlorine and
[17s\allowbreak12p\allowbreak5d]/(4s\allowbreak3p\allowbreak1d) contraction for sulfur.~\cite{Roos2005_6575, Roos2004_2851, Widmark1990} The active space for 
all Heme B derivatives and for all spin configurations, has been chosen to consist of three 2p orbitals in RAS1 and the
five 3d orbitals in RAS2 to describe dipole allowed $2p \rightarrow 3d$ electronic transitions. Further, one hole was allowed in RAS1 and 5 or 6 electrons in
RAS2, so that the total number of active electrons is 13
(cf. Fig.~\ref{fig:structure}). 

The spin-orbit coupling (SOC)  was included within the LS-coupling scheme and using the atomic mean-field integral approximation as implemented in the Molcas 8.0 program package.~\cite{Molcas, malmqvist:2002} 
At this point, the calculation has been restricted to sextet (\(S=5/2\)) corresponding to the ground state spin and quartet (\(S=3/2\)) spin configurations according to the spin-orbit selection rule \(\Delta S= 0, \pm 1\).
\cc{This resulted in 16 sextet and 174 quartet spin-free states or 792 spin-orbit coupled states, with 102 valence states covering energy range up to 7.3\,eV and 690 core states with energies between 710.6 and 732.9\,eV.
To cover this part of the spectrum within a supermolecule-type dimer calculation it would be required to include 61 and 34265 sextet and quartet spin-free states, respectively, or 137426 SOC states. This goes far beyond present computational capabilities.}
Scalar relativistic effects have been taken into account using the Douglas-Kroll-Hess transformation.~\cite{douglas:1974,hess:1986} 

Based on the local electronic wave functions, dimer states have been obtained as outlined
in Sec.~\ref{sec:theory}. In the results presented below we will consider the cases of
one- and two-particle approximations (OPA and TPA) for the sake of comparison. The distance of 7 a.u. between monomers was chosen as representative on the basis of molecular dynamics simulations~\cite{Villiers_2006} of this very non-rigid system, with heme ligand planes being parallel to each other and axial ligands pointing to the outside of the dimer. To study the effect of mutual orientation, three geometrical configurations have been considered: \(0^{\circ}\) rotated (COOH groups of both monomers are on one side of the dimer), \(90^{\circ}\) rotated, and \(180^{\circ}\) rotated (COOH groups are on the opposite sides).

Spectra have been calculated according to Eqs.~\eqref{eq:SXAS} - \eqref{eq:SPFY}.
The broadening is determined by the finite life times of the intermediate states as well as
through a Gaussian lineshape of the excitation pulse. 
In order to fit experimental conditions, the latter has been chosen to have a width of 0.25\,eV. 
Since even within the L$_3$-band the lifetimes may vary,~\cite{bokarev_reply} the  $\Gamma_n$ parameters in Eq.~\eqref{eq:SRIXS} have been assumed to be 0.09\,eV for states below 709.2\,eV, 0.26\,eV between 709.2 and 711.6\,eV, 0.43\,eV between 711.6 and 719.2\,eV, and 0.61\,eV above 719.2\,eV \cc{as a best fit to the experimental data}. 
The hemin monomer has a sextet ground state, where the degeneracy of the six components is slightly lifted due to SOC and thus the states are split into three Kramers doublets.
For the monomer spectra calculations, Eqs.~\eqref{eq:SXAS} and \eqref{eq:SRIXS}, these six  ground states are populated according to a Boltzmann distribution at a temperature of 300\,K. For the case of the dimer, 36 combinations of the ground states are possible; here for simplicity  we considered only first four of them resulting from lowest Kramers doublet of both monomers.
This effectively corresponds to very low temperatures. For the comparison, the same Kramers pair was used for the monomer calculations as well.

For the PFY spectra, the frequency interval  $[E_{\rm min}:E_{\rm max}]$ has
been chosen to range from 695\,eV to 735\,eV, corresponding to the
\(3d\rightarrow 2p\) radiative decay channel.
An absolute energetic shift of -2.8~eV has been applied to all theoretical spectra to ease the comparison with experiments.
%
\section{Results and Discussion}
\label{sec:res}
\subsection{Monomer Spectra}
\begin{figure}[t]
  \includegraphics{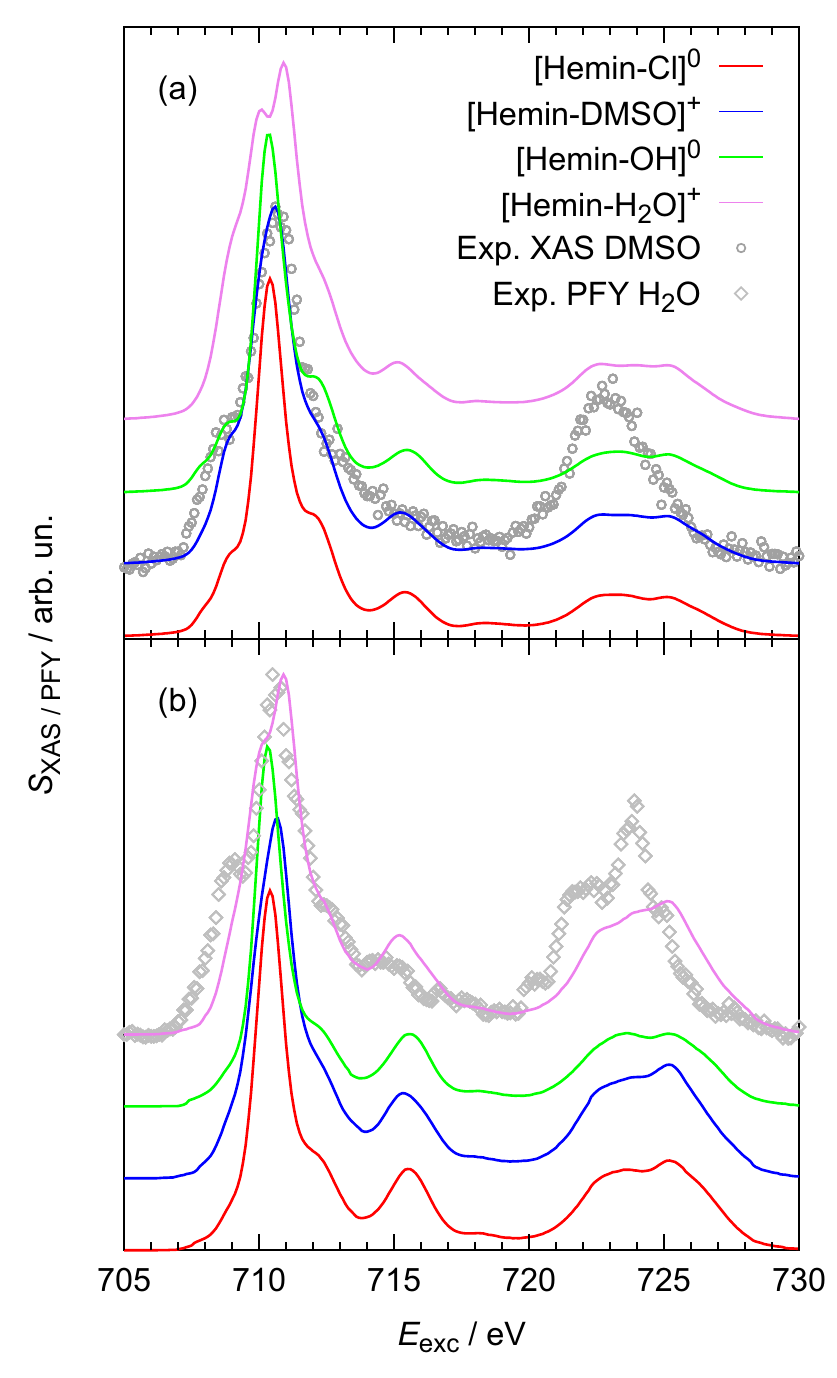}
  \caption{X-ray spectra for Heme-B with different ligands: (a) XAS and (b)
    PFY as compared to experimental data.~\cite{atak14_9938,golnak15_3058} All calculated
    spectra are normalized to the maximum of the experimental spectrum.}\label{fig_monomer_solvent}
\end{figure}

Before addressing effects of the excitonic coupling in the hemin dimer, the change of the XAS, PFY and RIXS spectra due to the different solvents need to be quantified for the case of the monomer.  The following discussion in this section is based on the assumption that there are different species in solution resulting from an exchange of the chloride ligand with solvent molecules.
Specifically, spectra have been calculated for various axial ligands: Cl$^-$ (original hemin), H$_2$O and DMSO, which have been used as solvents for experimental measurements \cite{atak14_9938,golnak15_3058},
and OH$^-$ that can be formed in aqueous solution due to hydrolysis ($\rm{Fe}^{3+}+\rm{H}_2\rm{O}\rightarrow \rm{FeOH}^{2+}+\rm{H}^+$). 
The XAS and PFY spectra of compounds with these ligands can be found in Fig.~\ref{fig_monomer_solvent}.

The XAS (Fig.~\ref{fig_monomer_solvent}(a)) shows a distinct sensitivity in the L\(_3\)-edge with respect to the ligand's nature.
For \HOH \, a splitting of this peak in two components at 710.0\,eV and 710.9\,eV can be observed and for \DMSO \, the lower energy feature appears as a shoulder at 710.1\,eV.  In contrast, for \Cl \, and \OH \, there is a single main L\(_3\)-peak only. 
In comparison to XAS, the PFY spectra (Fig.~\ref{fig_monomer_solvent}(b)) are less sensitive to the nature
of ligand, but show an intensity enhancement of all features for excitation energies above
716\,eV. The differences between XAS and PFY are caused by inelastic features that
become more important for higher excitation energies. An explanation for this fact will be
given after the interpretation of 1D-RIXS spectra (see Fig.~\ref{fig_monomer_h2o_slice}) below.

Due to the multi-configurational nature of the core-excited states, the excited 2p electron is mostly evenly distributed over 3d orbitals. However, the most prominent transitions in XAS are due to 2p $\rightarrow \rm{d}_{\rm{x^2-y^2}}$ and d$_{\rm{z}^2}$ excitations.  Interestingly, the splitting in XAS of \HOH is due to the energetic lowering of 2p$\rightarrow \rm{d}_{\rm{z^2}}$ with respect to 2p $\rightarrow \rm{d}_{\rm{x^2-y^2}}$ upon change of axial ligand from Cl$^-$ to H$_2$O.

\begin{figure}[t]
  \includegraphics{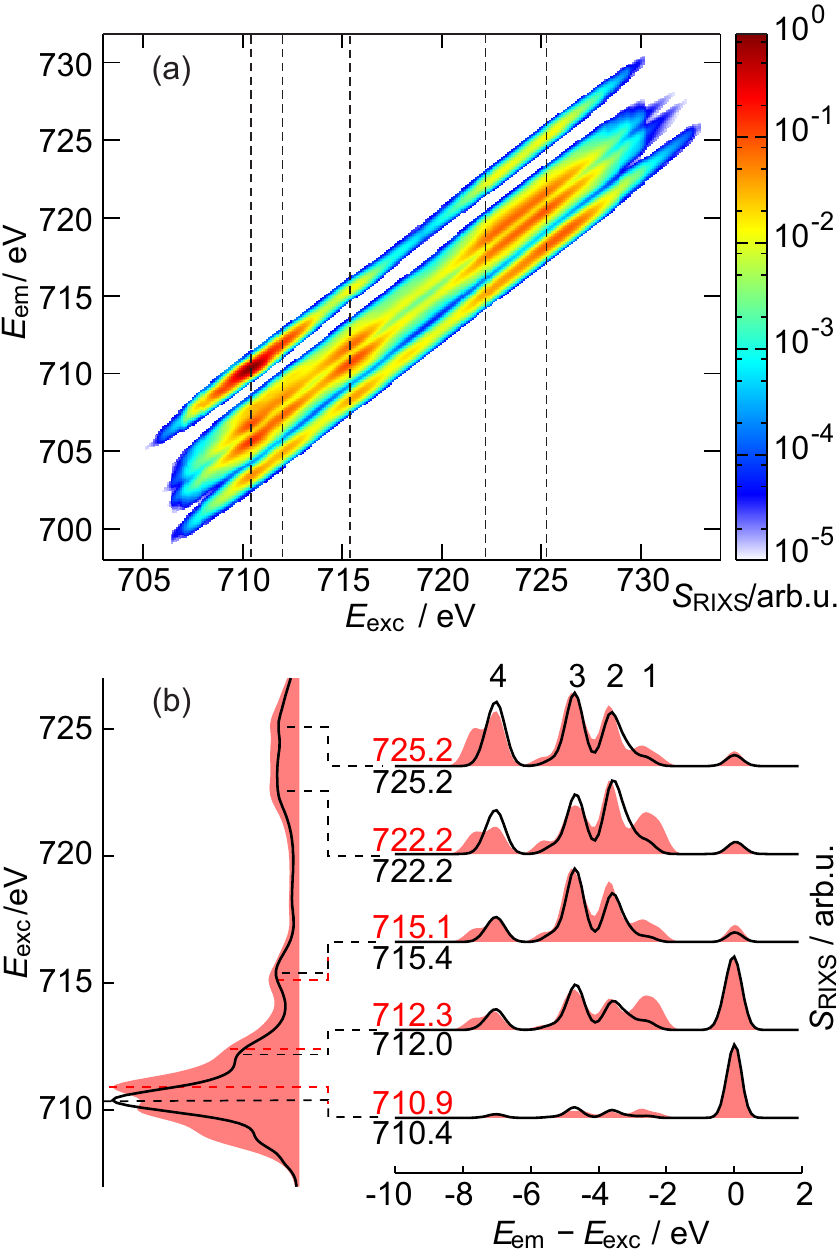}
  \caption{(a) 2D RIXS spectrum of \Cl; (b) Left panel: normalized XAS for \Cl (black) and \HOH (red filled curves). Right panel: normalized 1D-RIXS spectrum for selected excitation
    energies that are specified by the dashed lines and numbers. The peaks labeled 1-4 are discussed in the text.}
  \label{fig_monomer_h2o_slice}
\end{figure}

2D-RIXS spectra have been obtained for all Heme B derivatives. Exemplarily, Fig.~\ref{fig_monomer_h2o_slice}(a) shows the 2D-RIXS spectrum for the \Cl \, case.  However, since the analysis of 2D spectra is rather difficult and the differences between the species are not very pronounced in the 2D presentations, one-dimensional cuts of the RIXS spectra will be analyzed for \Cl \, and \HOH.  Five excitation energies, belonging to distinct spectroscopic features, were selected as shown in the left panel of Fig.~\ref{fig_monomer_h2o_slice}(b). 

The 1D-RIXS spectra show a prominent elastic peak for lower excitation energies, whose intensity decreases upon increasing the excitation
energy. This behavior can be rationalized as follows. Below 715\,eV  core-excited sextet states are dominating the
spectrum. Due to the spin selection rules, the preferred emission is to 
the sextet ground state, thus yielding an intense elastic peak. For excitation energies
above 720\,eV, the core-excited states are mostly of quartet type. Here, the most intense emission is the relaxation from
a core-excited state to a valence-excited state with a high quartet contribution. The
elastic peak corresponds to a relaxation to the (sextet) ground state, which is spin
forbidden and therefore less intense than the inelastic features.

The 1D-RIXS spectra for \Cl \, and \HOH \, differ mostly in the inelastic peaks, whereas the elastic peak has a comparable intensity in both spectra. Among the inelastic features, the peaks with loss energies
\(E_{\rm em}-E_{\rm exc}\)
of -2.5\,eV, -3.6\,eV, -4.7\,eV and -7.0\,eV are most prominent and labeled by
1-4 in Fig.~\ref{fig_monomer_h2o_slice}. 
All inelastic features in the RIXS spectra are due to the formally spin-forbidden transitions enabled by the strong SOC in the intermediate state. Despite the pronounced multi-configurational character the inelastic bands can be roughly assigned to the refill of core hole by the electrons from the following orbitals:
1) d$_{\rm{x^2-y^2}}$; 
2) d$_{\rm{z}^2}$;
3) d$_{\rm{xz}}$ and d$_{\rm{yz}}$;
4) d$_{\rm{xy}}$.
Interestingly, although  d$_{\rm{x^2-y^2}}$ and d$_{\rm{xy}}$ orbitals are not directly affected by axial ligand, the largest differences between \Cl and \HOH correspond to bands 1 and 4. Summarizing, RIXS spectra as well as XAS show a prominent sensitivity to the substitution of the axial ligands, with the largest changes being observed for \HOH \, case.
\begin{figure*}[th]
  \includegraphics{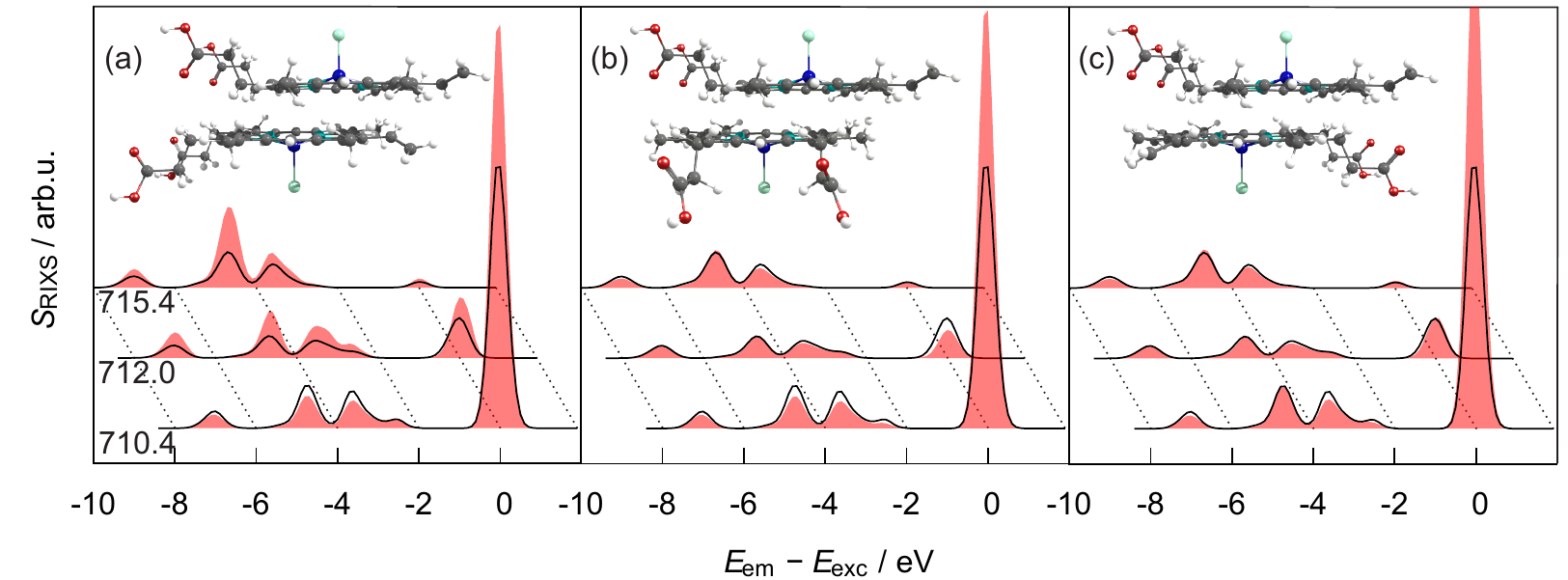}
  \caption{1D-RIXS spectra (not normalized) of the \Cl \, dimer (red filled curves) calculated with the TPA basis and different orientations of the COOH groups (a) \(0^{\circ}\), (b) \(90^{\circ}\), and (c) \(180^{\circ}\) for three excitation energies. The monomer spectrum is shown for comparison as well ($\times 2$, black lines).}
\label{extended_RIXS}
\end{figure*}

\subsection{Dimer Spectra}
Below we will focus mainly on the L$_3$-edge since it is more structured and less subject to the lifetime broadening than the L$_2$-edge.
The OPA does not show any notable difference between monomer and dimer 
spectra both for XAS and RIXS for all three orientations of the molecules. The reason for the minute differences is that the coupling of transition densities is rather weak for core excitations as compared to valence ones.
This can be attributed to the intensities of the metal \(2p \rightarrow 3d\) and  \(3d\rightarrow 3d\) transitions relevant for L-edge X-ray spectra, that are lower (due to smaller radial overlap and dipole selection rules) than those of the \(\pi \rightarrow \pi^*\) and \(n\rightarrow \pi^*\) transitions usually discussed in the case of organic dyes.
Moreover, the OPA exciton basis by construction should be appropriate only for the first-order XAS spectra (one-photon transitions). 
For the second-order RIXS it is natural to take the $|a_1c_2 \rangle$ ($a_1=g_1,v_1$) type of basis functions into account, since one needs to describe the two-photon $g \rightarrow c \rightarrow a$ transitions which are additionally interfering with each other (see Eq. \eqref{eq:SRIXS}). Indeed for RIXS spectra,  in contrast to the OPA exciton basis the TPA basis predicts  aggregation effects that are up to one order of magnitude larger (not shown).
It should be noted, however, that even for TPA the form of XAS essentially does not change upon dimerization. Distinct fingerprints of dimerization can be seen in RIXS, evidencing that absorption spectroscopy should be less sensitive to aggregation than RIXS.
Therefore, in the following we discuss only TPA results for RIXS spectra.

As was mentioned in the Section~\ref{sec:compdet}, to reduce the size of the Hamiltonian matrix (Eq.~\eqref{eq:hmat1}) and the number of terms in the \cc{innermost}  sum in Eq.~\eqref{eq:SRIXS} we have limited the basis to those states $| c \rangle$ which are within a $\pm$1.25~eV energy window around the prominent absorption features. 
\cc{This allowed to reduce the rank of Hamiltonian block to be diagonalized from 140964 to about 22000 (depending on the excitation energy).}
Such an approximation is justified by the finite width of the excitation pulse as well as the fast Lorentzian decay of interference terms in Eq.~\eqref{eq:SRIXS} with the energy separation between radiative channels. 
Further, recall that only four degenerate initial states were taken into account for the dimer case. For the purposes of comparison, the corresponding monomer spectra include only two degenerate initial states. Note that as far as the monomer is concerned the differences between two and six initial states are small as compared to the dimerization effects.

The 1D-RIXS spectra  in Fig.~\ref{extended_RIXS} show notable differences between the monomer and dimer cases: a prominent increase of intensities of elastic bands (especially for the \(180^{\circ}\) orientation) is observed, whereas the energy shifts are not larger than 0.1~eV. The spectra demonstrate a distinct orientation dependence, with the largest differences from the monomer case being observed for $0^{\circ}$ and the smallest for $180^{\circ}$. For $0^{\circ}$, the overall intensity mostly increases, for 90$^{\circ}$ slightly decreases, and for 180$^{\circ}$ stays almost intact. However, the high density of states hinders a detailled analysis in terms of, e.g, orientations of transition dipole moments. Qualitatively, one can say that apart from the elastic features, the largest deviations are observed for d-orbitals having an out-of-plane z-component (d$_{\rm{xz}}$, d$_{\rm{yz}}$, and d$_{\rm{z^2}}$). The "in-plane" transitions are affected only for the $0^{\circ}$ orientation. 

To summarize, although the transition moments for the monomers are quite small if compared to valence excitations of organic dyes, there is an effect of dimerization that can be seen in the RIXS spectra. Moreover, there appears to be a pronounced dependence on conformation. Given the fact that in solution the dimer system is floppy and interconversion between conformations is rather likely,~\cite{Villiers_2006} a direct comparison with experiment would  require, e.g., a computationally demanding molecular dynamics based sampling of spectra.

\section{Conclusions}
\label{sec:concl}
%
The Frenkel exciton model is usually applied to describe aggregation effects on spectra as well as excitation energy transfer in molecular aggregates.\cite{schroter15_1}  In the present contribution its basic idea has been adopted to the computation of the core-level spectra of multi-center transition metal compounds. Thereby, the RASSCF-based protocol for treating multiconfigurational and spin-orbit coupling effects has been extended to multi-center systems, which are not accessible by the standard protocol due to computational limitations. 
While in molecular aggregates only valence excitations are of relevance and often the treatment can be reduced to monomeric two level systems, the description of core-level spectra of transition metals requires to take into account a large number of possible transitions.
This renders the interpretation, e.g., in terms of a few coupled transition dipoles to become essentially impossible. Further, in standard exciton theory one usually classifies the collective excitations as zero-, one-, two-exciton states etc. This is particularly useful in the context of (non)linear spectroscopy.\cite{mukamel95_} In the present case, however, such a classification is not very useful due to the multitude of possible excitations. Still, different approximations derive from the used aggregate basis. Here, we discussed the one- and two-particle basis, the latter being exact for the dimer, but an approximation for a larger aggregate. It turned out that, similar to  standard exciton theory, the one-particle basis is suitable for describing one-photon processes like XAS only. The two-photon RIXS requires to take into account two-particle excitations.

In general, due to the  quite small transition dipole moments for the 2p$\rightarrow$3d excitations, if compared to the valence \(\pi \rightarrow \pi^*\) and \(n\rightarrow \pi^*\) transitions of organic dyes, the effect of aggregation on XAS spectra will be rather small. As far as the RIXS spectrum is concerned, Eq.~\eqref{eq:SRIXS} points to a dependence on the electronic polarizability. Here, the intermonomer contributions will determine the aggregation effect on the spectrum.

The developed protocol was applied to the hemin system forming dimers in water solution, while staying monomeric in other polar solvents.
Remarkably, ligand coordination in various solvents has been shown to have a pronounced influence for XAS and in particular  RIXS spectra. This is an important result, since solvent effects have not been considered in the previous experimental studies of hemin X-ray spectra.~\cite{atak14_9938,golnak15_3058,golnak15_29000}
In present work it was found that coordination and aggregation effects on RIXS spectra are of similar magnitude. This could make the unequivocal assignment of  aggregation induced features difficult.
However, in this proof-of-principle study a direct comparison with experiment was not attempted. First, due to construction of the active space  the $\pi-\pi$ stacking effect, discussed e.g. in Ref. \citenum{golnak15_3058}, was not taken into account. 
Second, it was found that the aggregation induced changes in the RIXS spectrum are depending on the mutual orientation of the monomers in the dimer. Due to the flexibility of the dimer structure in solution a more accurate description would require a rather time consuming sampling of different conformations, i.e. by combining molecular dynamics with the present RIXS calculation.

The present Frenkel exciton-like approach to the X-ray spectroscopy of multi-center systems should be particularly suitable to describe coupled \cc{highly-local} core excitations of weakly bound van der Waals  complexes. To include situations with more extended electron densities, the dipole approximation has to be replaced by a more accurate calculation of transition densities. This could be achieved using standard tools for integration of Gaussian or Slater-type orbitals. For situations where covalent bond formation is of importance or where the monomers are ferromagnetically and antiferromagnetically coupled, taking into \cc{account} the exchange contribution would be mandatory. Finally, the present approach has been developed for a molecular dimer only. However, the extension to larger aggregates is straightforward although eventually bound to the applicability of a certain $n$-particle basis, with $n$ being small enough to accommodate current computational resources.

The Frenkel exciton approach has been extensively used in the context of nonlinear spectroscopy and dynamics of dye aggregates. The present adaption to the X-ray regime in principle facilitates similar investigation for core-level excitations. Thus, upcoming ultrafast spectroscopic techniques in the X-ray regime \cite{zhang14_558,zhang15_273} could be a target for future advancement of the present approach.

\begin{acknowledgments}
  We acknowledge financial support by the Deanship of Scientific Research (DSR), King  Abdulaziz University, Jeddah, (grant No.\ D-003-435) and the Deutsche Forschungsgemeinschaft (grant No.\ KU952/10-1).
\end{acknowledgments}

\end{document}